\documentclass[dvipdfmx]{PoS}
\usepackage{amsmath}
\usepackage{subfigure} 
\usepackage{cite}
\allowdisplaybreaks[1]

\newcommand{\td}{{\rm{d}}}
\newcommand{\e}{{\rm{e}}}

\title{
\begin{picture}(0,0)(0,0)%
   \put(300,75){\makebox(0,0)[l]{\textnormal
{\normalsize OU-HET-922, KEK-CP-352
}
}}%
\end{picture}%
Current correlators in the coordinate space at short distances}

\ShortTitle{Current correlators in the coordinate space at short distances}

\author{JLQCD Collaboration:}

\author{\vspace{-5mm}\speaker{Masaaki Tomii}\\
        Physics Department, Columbia University, New York 10027, USA
        \\
        E-mail: \email{mt3164@columbia.edu}}
\author{Hidenori Fukaya\\
        Department of Physics, Osaka University, Toyonaka 560-0043, Japan}
\author{Shoji Hashimoto and Takashi Kaneko\\
        KEK Theory Center, Institute of Particle and Nuclear Studies, High Energy Accelerator Organization (KEK), Tsukuba 305-0801, Japan\\
        Department of Particle and Nuclear Physics, SOKENDAI (The Graduate University for Advanced Studies),
        Tsukuba 305-0801, Japan}

\abstract{
We calculate the vector and axial-vector current correlators in the coordinate space
and compare them with the experimental information obtained through the spectral
functions of hadronic $\tau$ decays measured by ALEPH.
Lattice data are obtained with 2+1 M\"obius domain-wall fermions at three lattice
spacings 0.044, 0.055 and 0.080 fm and the continuum limit is taken.
The correlators calculated on the lattice after extrapolating to the physical point
agree with those converted from the ALEPH data.
}

\FullConference{34th annual International Symposium on Lattice Field Theory\\
		24-30 July 2016\\
		University of Southampton, UK}

\begin{document}

\section{Introduction}

Current correlators provide a rich source of information on the QCD vacuum.
Their characteristics vary depending on the distance between the currents.
Correlators at short distances behave perturbatively providing the information on
the strong coupling constant, while those at long distances are saturated by the
ground state reflecting the individual mass spectrum and decay constant.
In the middle distances between perturbative and non-perturbative regimes,
on the other hand, perturbative approaches and low-energy effective theories
are no longer suitable to analyze correlators.

Lattice calculation is useful to calculate current correlators at any distances.
So far, lattice calculation of current correlators has been aimed mainly at long
distances to extract hadron masses and decay constants and shown the agreement
with experiments.
This agreement supports the consistency between experiments and QCD at
low energies, where contributions of excited states are missing and
only a part of QCD can be seen.
Comparison of lattice correlators at short and middle distances with experimental
observables may provide a test of the consistency between experiments and QCD at
higher energies corresponding to the excited states.

The vector and axial-vector current correlators are useful for such analysis because
they can be compared to the experimental observable in hadronic $\tau$ decays through
the dispersion relation.
The early ALEPH data \cite{Barate:1997hv,Barate:1998uf} of the $\tau$ decay
experiment were converted to the vector and axial-vector correlators \cite{Schafer:2000rv}
and the result was used to test the consistency with a quenched lattice calculation
at a single lattice spacing \cite{DeGrand:2001tm}.
%With unquenched lattice calculations and updated experimental data
%\cite{Davier:2013sfa} after these works, we have a good opportunity to revisit this subject.

With unquenched lattice simulations and updated ALEPH data
\cite{Davier:2013sfa}, we perform a more realistic calculation.
% as a part of our
%studies on current correlators at short and middle distances, which have determined
%renormalization factors of quark currents and the chiral condensate.
%In this work, we employ $2+1$-flavor M\"obius domain-wall fermions with three-step
%stout-link smearing.
%Numerical simulations are carried out on $32^3\times64$, $48^3\times96$ and
%$64^3\times128$ lattices at
%the lattice spacings $a = 0.080, 0.055$ and 0.044~fm, respectively.
Numerical simulations are carried out using $2+1$-flavor M\"obius domain-wall fermions
on $32^3\times64$, $48^3\times96$ and $64^3\times128$ lattices at
the lattice spacings $a = 0.080, 0.055$ and 0.044~fm, respectively.
Pion masses on our ensembles are 500, 400, 300 and 230~MeV.
We show the extrapolation to the physical pion mass and continuum limits agrees
with the ALEPH data.
%This analysis is done as a part of our studies on short- and middle-distance current
%correlators, in which we have determined renormalization factors of quark currents
%\cite{Tomii:2016xiv} and the chiral condensate \cite{Tomii:2015exs}.

This analysis is done as a part of our studies on current correlators at short and middle
distances to investigate what kinds of information of QCD we can extract from them.
In these studies, we have determined renormalization factors of quark currents
\cite{Tomii:2016xiv} and the chiral condensate \cite{Tomii:2015exs}.

%We use the local vector and axial-vector currents, whose renormalization
%factor is determined \cite{Tomii:2016xiv} by matching the correlators in perturbative
%region to those in continuum perturbation theory, which is known to the order of
%$\alpha_s^4$ \cite{Chetyrkin:2010dx}.

\section{Current correlators}

We calculate the vector and axial-vector current correlators
\begin{equation}
\Pi_V (x) = \sum_\mu \langle V_\mu(x)V_\mu(0)^\dag\rangle,
\hspace{10mm}
\Pi_A (x) = \sum_\mu \langle A_\mu(x)A_\mu(0)^\dag\rangle
-\Pi_A^{(0)}(x),
\end{equation}
where the vector and axial-vector currents $V(x)$ and $A(x)$ are iso-triplet.
Since the spectral functions obtained through hadronic $\tau$ decays are related
to the spin-1 contribution of the correlators,
we subtract from the axial-vector channel the spin-0 part estimated
as the contribution of the ground state pion
$\Pi_A^{(0)}(x) \simeq z_0M_\pi^2 K_1(M_\pi|x|)/2\pi^2|x|$, where $K_1$ stands for the
modified Bessel function and $z_0$ and $M_\pi$ are extracted from the zero-momentum
correlator of the axial-vector current,
$\int\td^3x\langle A_4(x)A_4(0)^\dag\rangle \rightarrow z_0\e^{-M_\pi x_4}$.
The vector channel does not need any modification due to the absence of the
spin-zero part in the isospin limit.
We subtract the discretization effect at the mean field level.
The detail is explained in \cite{Tomii:2016xiv}.

These correlators are related to the experimentally measured spectral functions
$\rho_{V/A}(s)$ through the dispersion relation \cite{Schafer:2000rv}
\begin{equation}
\Pi_{V/A}(x) =
{3\over8\pi^4}\int_0^\infty\td s\ s^{3/2}\rho_{V/A}(s){K_1(\sqrt{s}|x|)\over|x|}.
\label{eq:disp_rel_spin1}
\end{equation}
The hadronic $\tau$ decay experiment provides the spectral functions $\rho_{V/A}(s)$
at invariant masses smaller than the $\tau$ lepton mass, $s < m_\tau^2$.
We use the latest ALEPH data $\rho_{V/A}(s)$ of the $\tau$ decay experiment
\cite{Davier:2013sfa} in this region.
In the region $s > m_\tau^2$, we calculate $\rho_{V/A}(s)$ using perturbation theory
available to order $\alpha_s^4$ \cite{Baikov:2008jh} and a model \cite{Cata:2008ye}
to take account of the violation of the quark-hadron duality \cite{Shifman:1994yf,Blok:1997hs}.

\section{Lattice vs Experiment}

We compare correlators on the lattice with those obtained from the experimental
data for the sum and difference $R_{V\pm A}(x) = (\Pi_V(x)\pm\Pi_A(x))/2\Pi_0(x)$
normalized by the vector correlator in massless free theory $\Pi_0(x)$.
The $V+A$ channel contains the perturbative contribution, which dominates
at short distances.
The $V-A$ channel consists only of the effects of the spontaneous chiral symmetry
breaking.

%=   Figure   ======================
\begin{figure}[t]
%\vspace{4.2mm}
\begin{center}
\includegraphics[width=100mm]{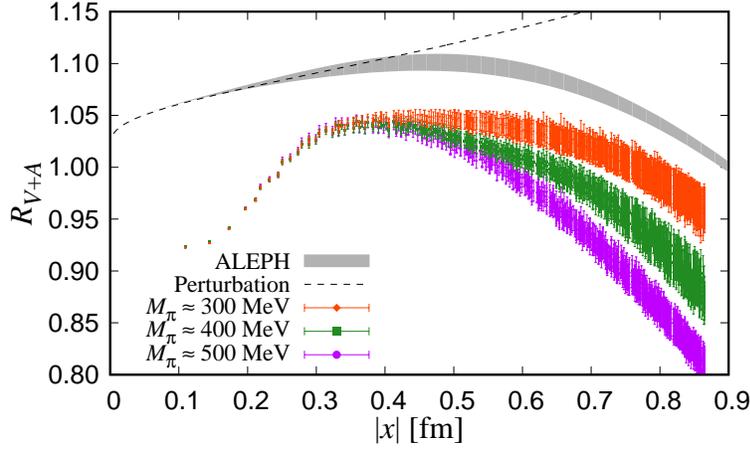}
\caption{
$R_{V+A}$ calculated on the ensembles at $a = 0.055$~fm
and the three pion masses $M_\pi \simeq 300$~MeV (diamonds), 400~MeV (squares) and
500~MeV (circles).
The prediction of massless perturbation theory (dashed curve) and the result from the
experiment (band) are also shown.
}
\label{fig:vpa_ms0.0180}
\end{center}
\end{figure}
%===============================

\begin{figure}[t]
%\vspace{4.2mm}
\begin{center}
\includegraphics[width=100mm]{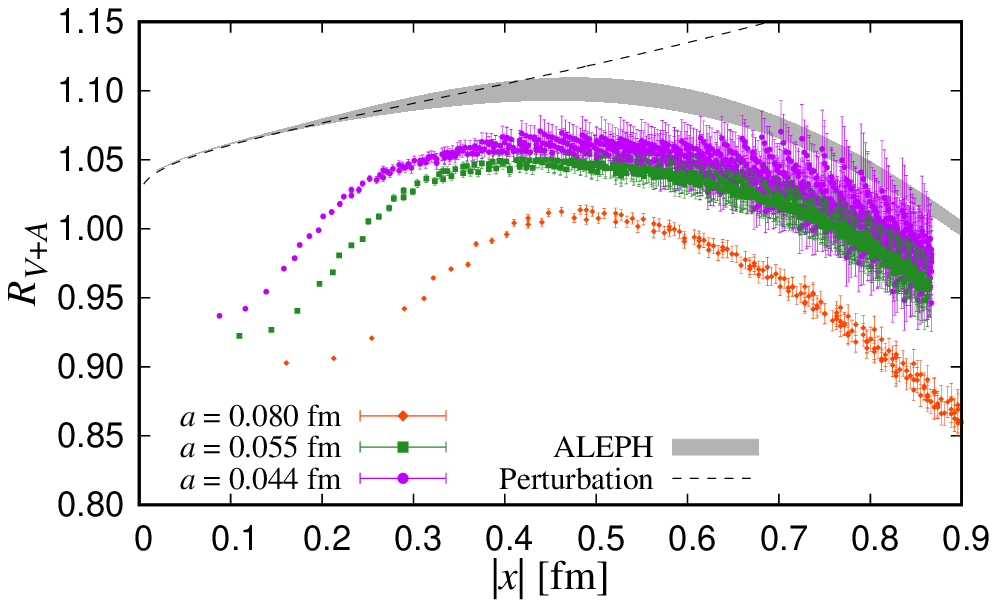}
\caption{
Same as Fig.~1 but calculated on the ensembles at
matched pion mass $M_\pi\simeq300$~MeV and different lattice spacings
$a=0.080$~fm (diamonds), 0.055~fm (squares) and 0.044~fm (circles).
}
\label{fig:vpa_3beta}
\end{center}
%\end{figure}
%===============================
%=   Figure   ======================
%\begin{figure}[t]
%\vspace{4.2mm}
\begin{center}
\includegraphics[width=100mm]{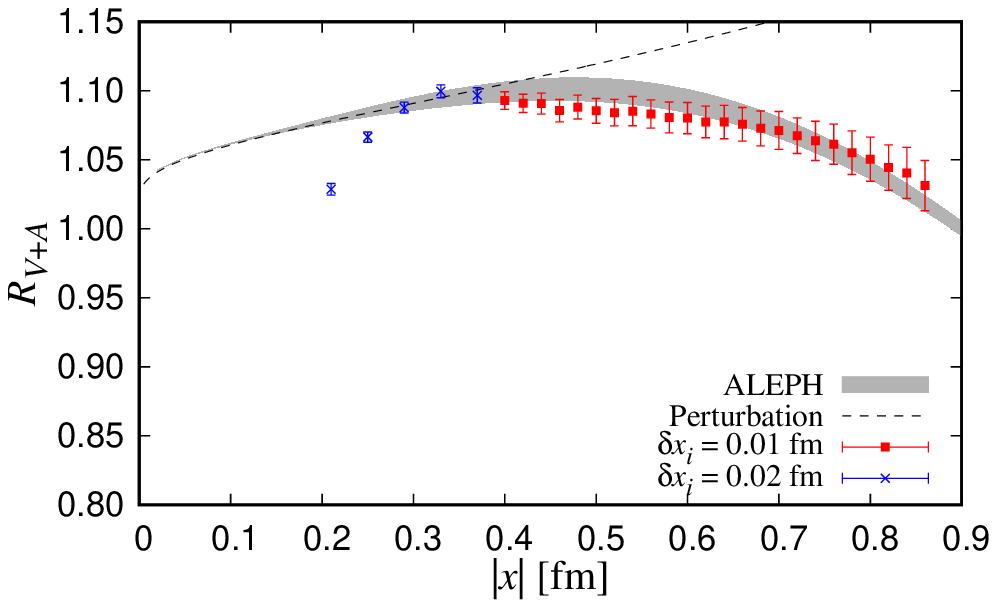}
\caption{
Extrapolation of $R_{V+A}$ to the physical point.
}
\label{fig:ALEPHvpa_ext}
\end{center}
\end{figure}

First, we focus on the $V+A$ channel.
Figure~\ref{fig:vpa_ms0.0180} shows the results of $R_{V+A}(x)$ calculated on
the ensembles at $a = 0.055$~fm and three different pion masses.
The results at smaller masses are closer to the experimental result (band),
implying that the chiral extrapolation to the physical pion mass makes them close
to the experimental data.
Figure~\ref{fig:vpa_3beta} shows the result at matched pion mass
$M_\pi \simeq 300$~MeV and at different lattice spacings.
There is a significant dependence on the lattice spacing, which can be
controlled by a term $\propto a^2$ at middle and long distances.
%The fact
%that the result at a smaller lattice spacing is closer to the experimental result
%implies the continuum limit is also expected to get close to the experiment.

%=   Figure   ======================
%===============================

We extrapolate these lattice results to the physical point, {\it i.e.} the continuum
limit $a\rightarrow0$ and physical pion mass limit
$M_\pi\rightarrow m_\pi\simeq140$~MeV as follows.
We divide the range of $|x|$ into $N$ bins,
\begin{equation}
B_i = [x_i-\delta x,x_i+\delta x],\ \ x_{i+1} = x_i+2\delta x,\ \ i = 1,2,\ldots, N,
\end{equation}
where $x_i$ and $\delta x$ are the center of the $i$th bin and one half of the
width of bins, respectively.
For each bin, we define $\overline R_{V+A}(a,M_\pi,x_i)$ as an average of
$R_{V+A}(x)$ over the lattice points falling in $B_i$.
We then perform a global fit for all ensembles using the fit function
\begin{equation}
\overline R_{V+A}(a,M_\pi;x_i)
= R_{V+ A}(0,m_\pi,x_i) + c_{m,i}(M_\pi^2-m_\pi^2)
+ c_{a,i}a^2,
\end{equation}
with three free parameters $R_{V+A}(0,m_\pi,x_i), c_{m,i}$ and $c_{a,i}$
for each $i$.
The first parameter $R_{V+A}(0,m_\pi,$ $x_i)$ corresponds to the extrapolated value.
The other parameters $c_{m,i}$ and $c_{a,i}$ are to control the dependences on
the pion mass and the lattice spacing, respectively.
Figure~\ref{fig:ALEPHvpa_ext} shows the result of the extrapolation.
Here, we take $\delta x = 0.01$~fm for $x_i\ge0.4$~fm and
$\delta x=0.02$~fm for $x_i<0.4$~fm.
%The consistency with the ALEPH data is seen in $|x| > 0.3$~fm.
After such an extrapolation, the lattice results are consistent with the experimental
data at $|x|\sim0.3$~fm and longer.
At even shorter distances, we find deviations from the perturbative and experimental
results, which suggest that the discretization effects beyond $O(a^2)$ are significant.
In fact, the coefficient $c_{a,i}$ rapidly increases toward the short-distance region.

%=   Figure   ======================
\begin{figure}[t]
%\vspace{4.2mm}
\begin{center}
\includegraphics[width=100mm]{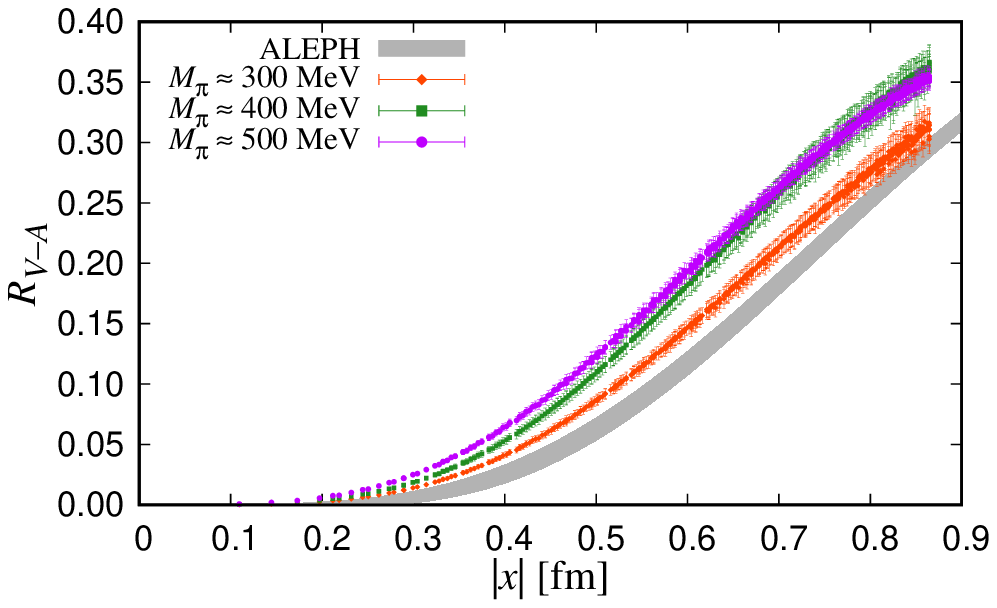}
\caption{
$R_{V-A}$ calculated on the same ensembles as in Fig.~1,
plotted with the experimental result (band).
}
\label{fig:ALEPHvma}
\end{center}
%\end{figure}
%===============================
%=   Figure   ======================
%\begin{figure}[t]
%\vspace{4.2mm}
\begin{center}
\includegraphics[width=100mm]{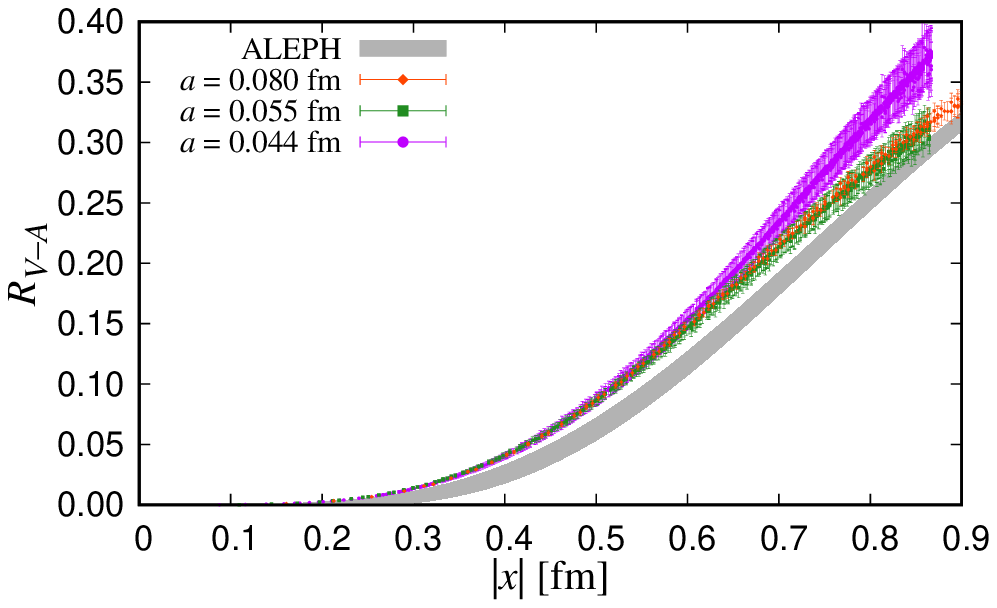}
\caption{
Same as Fig.~4 but calculated on the same ensembles as in
Fig.~2.
}
\label{fig:ALEPHvma_3beta}
\end{center}
\end{figure}
%===============================

%=   Figure   ======================
\begin{figure}[t]
%\vspace{4.2mm}
\begin{center}
\includegraphics[width=100mm]{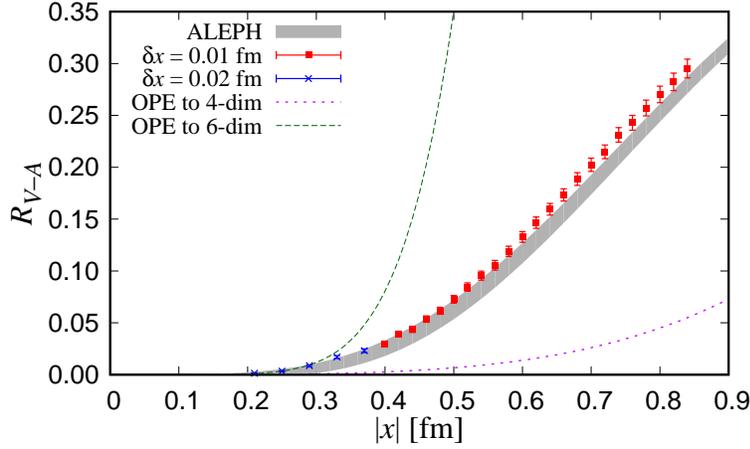}
\caption{
Extrapolation of $R_{V-A}$ to the physical point.
The experimental result (band) and the predictions of OPE including up to
four- (dotted curve) and six-dimensional (dashed curve) operators are
also plotted.
}
\label{fig:ALEPHvma_ext}
\end{center}
\end{figure}
%===============================

Next, we show the lattice results for the $V-A$ channel.
Figure~\ref{fig:ALEPHvma} shows the lattice data for $R_{V-A}(x)$ at $a=0.055$~fm and
three different pion masses.
$R_{V-A}(x)$ vanishes in the short-distance limit as it should be.
Actually it is guaranteed by the good chiral symmetry of M\"obius domain-wall fermions.
At short distances ($|x|\lesssim0.5$~fm), the dependence on the pion mass is clearly
seen and the results at smaller masses are closer to the experimental result.
In Fig.~\ref{fig:ALEPHvma_3beta}, which shows the data at the matched pion mass
$M_\pi \simeq 300$~MeV and three different lattice spacings, no significant dependence
on the lattice spacing is seen at least at short distances unlike the case of the $V+A$
channel.
One possible reason for this is that the most of discretization effect on correlators at
short distances is perturbative and cancelled for the $V-A$ channel.
Figure~\ref{fig:ALEPHvma_ext} shows the result of the extrapolation to the
physical point, which is done in the same manner as for the $V+A$ channel.
A good agreement with the ALEPH data is found.

%=   Figure   ======================
\begin{figure}[t]
%\vspace{4.2mm}
\begin{center}
\includegraphics[width=100mm]{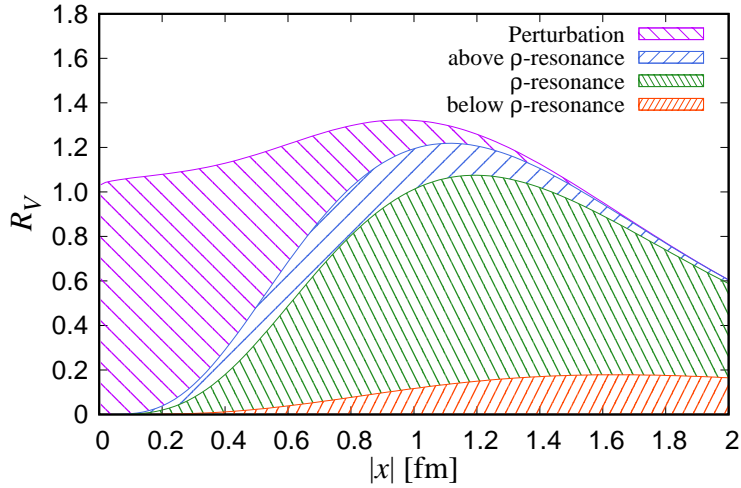}
\caption{
Decomposition of the vector correlators into the contributions of the spectral functions
in several regions of $s$.
The area indicated by ``Perturbation'' represents the contribution of the
spectral function in $s>2.7\rm~GeV^2$, which is calculated perturbatively.
The region $s\le2.7\rm~GeV^2$ is splitted
into the $\rho$ meson resonance
$(0.776 -0.150)^2{\rm~GeV}^2 < s < (0.776 + 0.150)^2~{\rm GeV}^2$
and above and below.
}
\label{fig:dominance}
\end{center}
\end{figure}
%===============================

We also investigate the region of $|x|$ where perturbative approaches give
reliable predictions.
Figure~\ref{fig:ALEPHvma_ext} also shows a rough prediction of the Operator
Product Expansion (OPE) \cite{Shifman:1978bx} truncated at dimension-four (dotted curve)
and dimension-six (dashed curve) operators.
The OPE truncated at the dimension-four underestimates the lattice
data already at 0.3~fm. The OPE including dimension-six operators
also disagrees with the lattice result in $|x|>0.3$~fm.
It indicates that the OPE of $R_{V-A}(x)$ is useful in the limited region $|x| < 0.3$~fm.
Figure~\ref{fig:dominance} shows the decomposition of
$R_V(x) = \Pi_V(x)/\Pi_0(x)$ from the experiment into the contributions of the
spectral function in several divided regions of $s$.
Since the non-perturbative contribution is substantial or even dominant at
$|x|\gtrsim0.5$~fm, the correlator calculated perturbatively in such a region may not
be precise.
The axial-vector channel also has a similar decomposition.

\section{Summary}

We calculate the vector and axial-vector current correlators in the short- and
middle-distance regions.
We show that the lattice results agree with the ALPEH data of hadronic $\tau$ decays
in $|x| > 0.3$~fm for the $V+A$ channel and $|x| > 0.2$~fm for the $V-A$ channel.

We also investigate the region where perturbative approaches are useful to describe
the correlators.
The OPE of the $V-A$ channel disagrees with the lattice result in $|x|\gtrsim0.3$~fm.
The individual channels of the correlators in $|x|\gtrsim0.5$~fm are not dominated
by the contributions from the perturbative regime.

\vspace{5mm}
Numerical simulations are performed on Hitachi SR 16000 and IBM System
Blue Gene Solution at KEK under a support of its Large Scale Simulation Program
(No.~15/16-09, 16/17-14).
We thank P. Boyle for the optimized code for BGQ.
This work is supported in part by the Grant-in-Aid of the Japanese Ministry of Education
(No.~25800147, 26247043, 26400259) and the Post-K supercomputer project through JICFuS.

\end{document}